# Ceres and the Terrestrial Planets Impact Cratering Record

Strom, R.G.[1], S. Marchi[2], and R. Malhotra[1]

[1] Lunar and Planetary Laboratory, The University of Arizona, Tucson, AZ. 85721

[2] Southwest Research Institute, Boulder, CO. 80302

## Abstract

Dwarf planet Ceres, the largest object in the Main Asteroid Belt, has a surface that exhibits a range of crater densities for a crater diameter range of 5-300 km. In all areas the shape of the craters' size-frequency distribution is very similar to those of the most ancient heavily cratered surfaces on the terrestrial planets. The most heavily cratered terrain on Ceres covers ~15% of its surface and has a crater density similar to the highest crater density on <1% of the lunar highlands. This region of higher crater density on Ceres probably records the high impact rate at early times and indicates that the other 85% of Ceres was partly resurfaced after the Late Heavy Bombardment (LHB) at ~4 Ga. The Ceres cratering record strongly indicates that the period of Late Heavy Bombardment originated from an impactor population whose size-frequency distribution resembles that of the Main Belt Asteroids.

## Introduction

Ceres is the largest object in the main asteroid belt, with an orbit near the center of the belt at a semi-major axis of 2.77 AU, a moderate eccentricity (e=0.076) and inclination of 10.6 degrees to the ecliptic. With a diameter of 945 km, it is much smaller than the Moon (3475 km) and only about the same size as the 950 km diameter lunar Orientale basin (Figure 1). The NASA Dawn mission obtained high-resolution images, among other measurements, of Ceres.

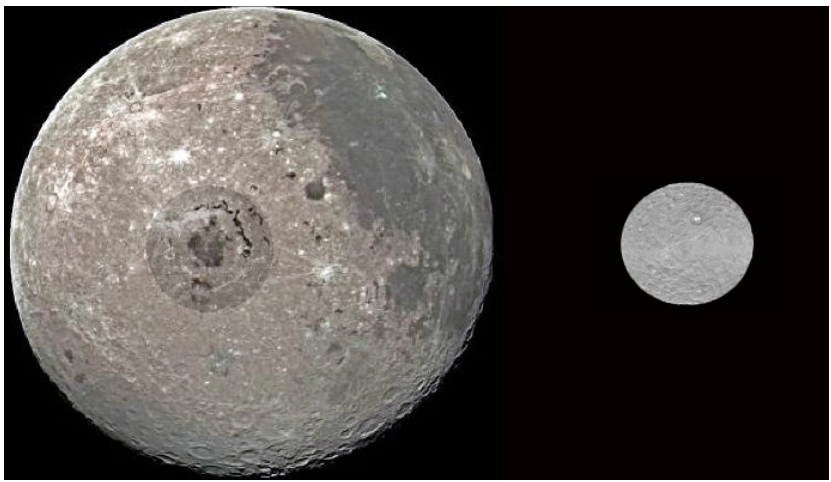

Figure 1. Size comparison of the Moon (3476 km dia.) and Ceres (945 km dia.). Ceres is about the same diameter as the lunar Orientale basin shown by the shaded circle.

A comparison of the cratering record on Ceres with the cratering record of the Moon and the terrestrial planets provides important new information on the origin of the impacting objects through time and the cratering process itself. Interpreting Ceres' cratering record is complicated by the evidence for cryovolcanic activity in the geologically recent past (Marchi, et al. 2016; Krohn, et al., 2016; Russell, et al. 2016; De Sanctis, et al. 2016; Ruesch, et al., 2016) involving hydrated salts resulting from global aqueous alteration of minerals on the surface, as well as by global relaxation of the surface at long-wavelengths (Fu, et al., 2017). A recent study shows that water must be present in the regolith in order to account for the water in Ceres' exosphere (Landis, et al. 2017).

## Terrestrial Planet Cratering Record

To facilitate comparison of Ceres' crater record with that of the Moon and terrestrial planets, this section provides a summary of the terrestrial planet cratering record with R plots of the crater and impactor size-frequency distributions taken from Strom, et al., 2015. The crater size-frequency distributions in the inner solar system indicate that the Moon and terrestrial planets were impacted by two primary populations of objects (Figure 2). Population 1 was due to the period of Late Heavy Bombardment (LHB) that began some time before ~4 Ga, peaked and declined rapidly over the next ~100 to 300 Myr (Tera et al. 1973, 1974, Ryder 1990) and possibly much more slowly from about 3.8–3.7 Ga to ~2 Ga (Bottke et al 2012, Morbidelli et al 2012, Marchi et al 2013). It produced heavily cratered surfaces with a complex, multi-sloped crater size-frequency distribution (Figure 2). Population 2 dominated since about 3.8–3.7 Ga. It produced a crater size distribution characterized by a differential –3 single-slope power law for crater diameters up to at least 100 km (Figure 2).

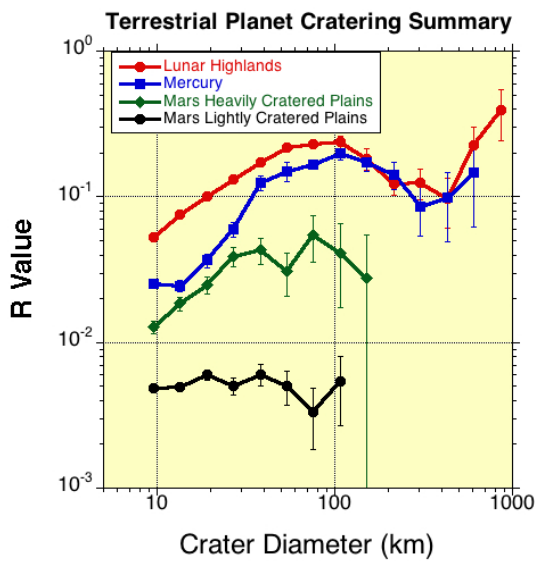

Figure 2. These *R* plots summarize the inner solar system cratering record for crater sizes in the range of about 10 km to about 1000 km. They show two distinctly different crater populations. The curves above an *R* value of about 0.01 have a complex shape characteristic of Population 1, and the lower curve (Mars Lightly Cratered Plains) has a nearly horizontal straight line shape characteristic of Population 2. Other lightly cratered regions of the Moon, Venus and Mercury have similar curves. The paucity of Mercury craters below about 35 km is due to intercrater plains formation. Also the upturn in the curve at 10 km diameter is due to secondaries. (See Strom et al., 2015 for additional details.)

A comparison of impactor sizes derived from the cratering record compared to the size distribution of main belt asteroids indicates that Population 1 had a size-frequency distribution that resembles that of the present-day MBAs, whereas the size-frequency distribution of Population 2 projectiles resembles that of the present-day near-Earth objects (see Figure 14 in Strom et al. 2015). Significantly, an observed systematic shift of the peak of the R plots of the crater SFDs found on the heavily cratered terrains of Mars, Moon and Mercury is consistent with that expected from the systematic difference in the average impact velocity of projectiles sourced from the main belt and hitting these three different bodies (see Figure 15 in Strom et al. 2015).

Strom et al. (2005) suggested the simplest interpretation is that Population 1 were asteroids from the ancient main asteroid belt delivered from their source region into terrestrial planet-crossing orbits, possibly by means of gravitational resonance sweeping during orbit migration of the giant outer planets. This caused the period of Late Heavy Bombardment (LHB). Similar results were obtained by comparing the crater SFD on most ancient lunar terrains with those somewhat younger of Nectarian-era (Marchi et al 2012). This scenario of the origin of the LHB impactors from the main asteroid belt is additionally supported by numerical simulations of planetesimal-driven late migration of the giant planets (Gomes et al. 2005; Bottke et al., 2012), as well as the signatures of that migration on the orbital distribution in the present-day asteroid belt (Minton & Malhotra 2009), although the precise timing of this event remains debated (Zellner 2017, Nesvorny et al. 2017). The mechanism of gravitational resonance sweeping would be insensitive to the size and mass of the asteroids, hence would produce projectiles having the same size distribution as that of the asteroid belt itself.

## The Ceres Cratering Record

The cratering record on Ceres in the size range 5-100 km has a crater size-frequency distribution that is very similar to the heavily cratered surfaces of the Moon, Mars, and Mercury. However, the crater density varies in certain areas of the asteroid. Figure 3 shows an area that has a very high density of craters while Figure 4 shows an area that is relatively lightly cratered. Figure 5 is R plots of heavily cratered (orange), lightly cratered (green) and total surface (blue) areas compared to the lunar highlands (red). A map of the counted craters in the heavily cratered terrain (total area 4.18 x $10^5$ km$^2$) and the lightly cratered terrain (total area 3.54 x $10^5$ km$^2$) as well as over the whole Ceres surface is shown in Figure 6; these crater counts have been published in Hiesinger et al. (2016) and Marchi et al. (2016).

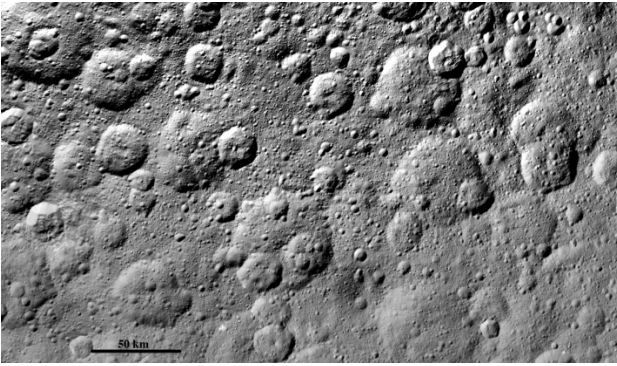

Figure 3. Photomosaic of a portion of the the heavily cratered area on Ceres. The region shown here is centered on 43° N and 315° E. The scale bar is 50 km.

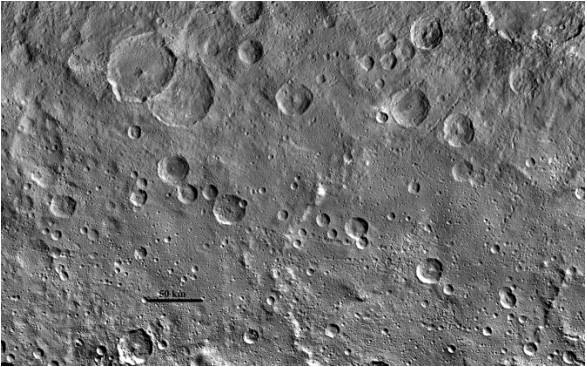

Figure 4. Photomosaic of a portion of the lightly cratered area on Ceres. The region shown here is centered on 0° N and 108° E. The large partial crater rim on the right has been almost obliterated by resurfacing. The scale bar is 50 km.

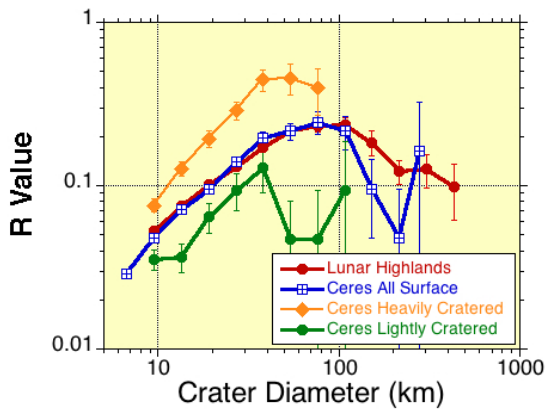

Figure 5. R plots of the lunar heavily cratered highlands (red) compared to the heavily cratered (orange) terrain, the entire surface (blue), and the lightly cratered (green) terrain of Ceres. See text for interpretations.

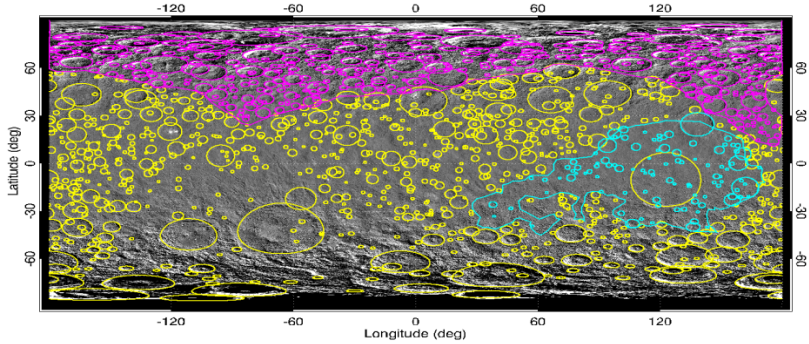

Figure 6. The craters counted in the "heavily cratered" and "lightly cratered" areas (for the R plots in Figure 5) are shown in magenta and cyan, respectively; craters counted over the rest of the Ceres surface are shown in yellow.

The total Ceres surface in the crater size range 20-100 km has a crater size-frequency distribution similar to Population 1 heavily cratered surfaces on the terrestrial planets. Therefore, the obvious interpretation is that the impacting objects have the same origin. There is one area on Ceres comprising 15% of the surface that has a very high crater density about twice the average lunar highlands crater density (see Figures 3 and 6). Figure 7 shows images for a comparison of Ceres' heavily cratered terrain and a portion of the lunar highlands near Tycho at the same scale. Visual inspection of the two images also shows that Ceres' heavily cratered terrain is indeed more heavily cratered than this part of the lunar highlands. It should also be noted that the lunar highlands exhibit a range of crater densities (Head et al 2011), with some regions exceeding the average, such as those most heavily cratered terrains discussed by Xiao and Werner (2015) and the Al-Khwarizmi King basin terrain discussed by Marchi et al (2012). The latter have an R value of ~ 0.3-0.4 at 70 km, similar to Ceres' high crater density; this can be compared to 0.23 for the average lunar highlands. This suggests that Ceres and the Moon had a similar impact history.

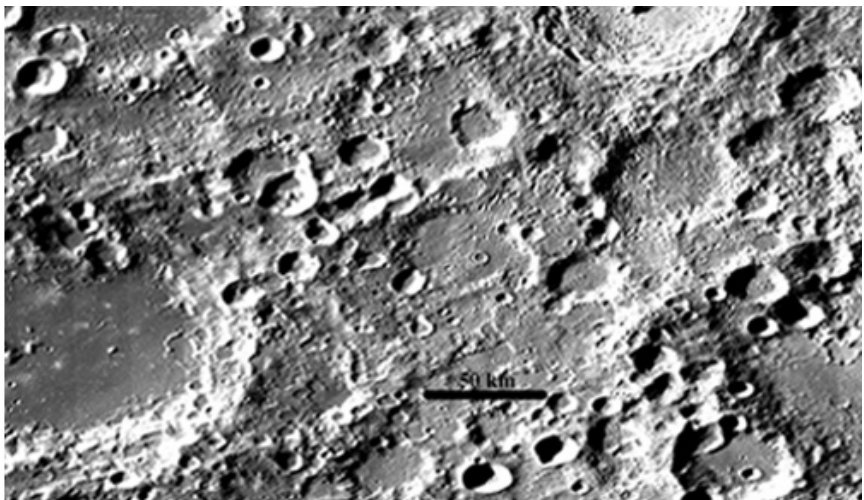

Lunar Highlands

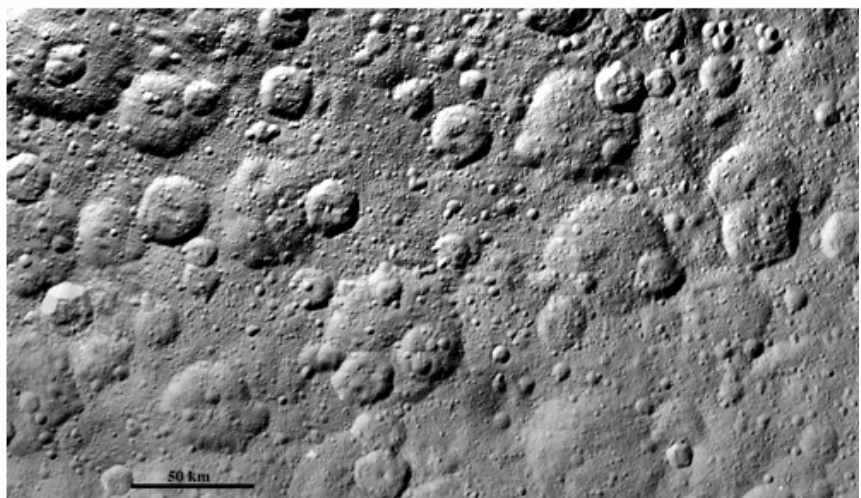

Ceres Heavily Cratered Terrain

Figure 7. The bottom image is a portion of the Ceres heavily cratered terrain and the top image is a portion the heavily cratered lunar highlands near the 86 km diameter Tycho crater at the top right of the image. The images are the same scale and the scale bars are 50 km. A visual comparison shows Ceres is more heavily cratered.

Richardson (2009) finds that saturation occurs between R values of 0.1 and 0.3 depending on the size frequency distribution. However, the slope of the Ceres heavily cratered curve at diameters less than about 35 km is steeper than the lunar highlands. This is probably due to saturation of the most heavily cratered terrain on the Moon. Richardson's simulations of saturation using the lunar highlands crater population show that at an R value of 0.4 the slope steepens (see Figure 13 in Richardson, 2009). Since the Ceres heavily cratered terrain has a similar shape and an R value of about 0.4 it is probably saturated. Both the Ceres and lunar most heavily cratered areas are close to the empirical saturation line. This reinforces prior arguments that the terrain used for the classical lunar highlands counts are not saturated. It is intriguing that two very different objects like the Moon and Ceres seem to have experienced similar impact and resurfacing histories.

On Ceres, the higher density is plausibly due to the continuous impact of asteroids from the Main Belt during and after the Late Heavy Bombardment. The cratering rate would have been higher within the asteroid belt during the LHB than in the inner solar system. During the LHB the whole surface probably became saturated. After the LHB ended about 3.7-3.8 Ga, 85% of Ceres was probably partially resurfaced by a combination of cryovolcanism and the largest impact craters that lowered the crater density to that observed for the other parts of the surface. Figure 8 shows two areas in the equatorial region of Ceres; in one area, some craters have been strongly degraded, and in the other are small plains areas where craters have probably been destroyed. The lightly cratered area shown in Figure 7 also shows highly degraded craters and plains that have obliterated craters resulting in a lower crater density. A similar resurfacing event probably occurred in the lunar highlands, but involving volcanism. In order to lower the crater density to that observed on Ceres, an average of about 55% of the craters would need to be erased. Although this may seem to be a very large amount, the lightest cratered area has only slightly lower crater density than the lunar highlands crater density. This would not be like the resurfacing of the Moon by mare formation. It would be much less intense over a long period of time.

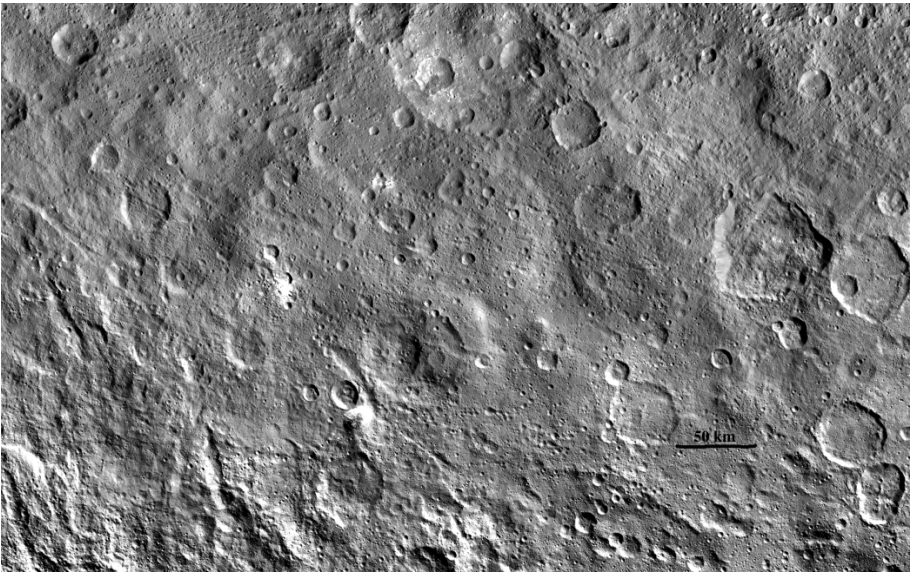

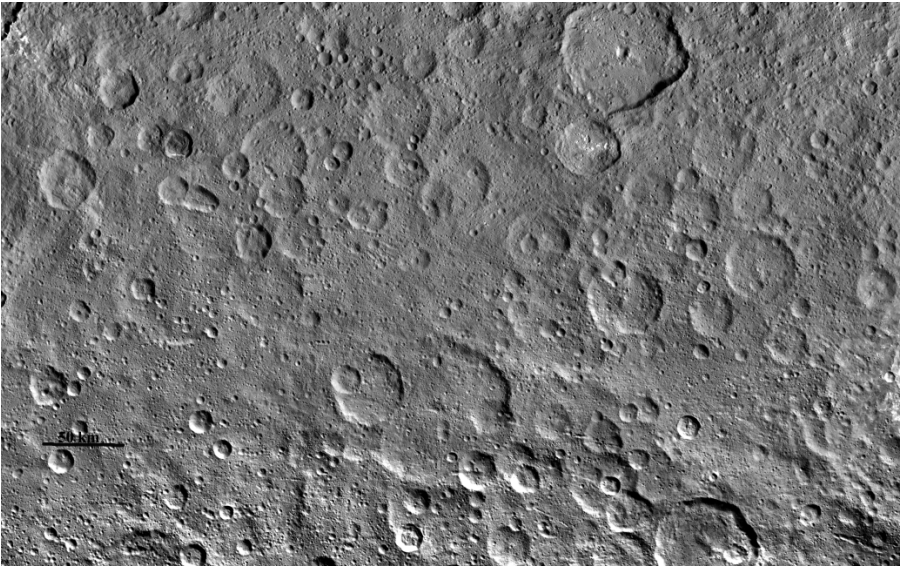

Figure 8. (Top) This region of Ceres is centered on the equator at 324 E. It shows smooth areas with few craters and the barely discernable rims or partial rims of craters. These areas have probably been partially resurfaced. (Bottom) This region of Ceres is also centered on the equator at 180 E. The center of the area shows smooth areas with few craters and highly degraded craters, some with missing rims. The scale bars are 50 km long.

## Origin of the Late Heavy Bombardment

Since Ceres is near the center of the asteroid belt its cratering record must be primarily due to impacts of asteroids. It has a very similar SFD as Population 1 on the Moon and terrestrial planets. Therefore, Population 1 is consistent with an origin from Main Belt Asteroids during the Late Heavy Bombardment. This is strengthened by the derivation of impactor diameters derived from the Melosh/Beyer (http://pirlwww.lpl.arizona.edu/~rbeyer/crater_p.html) equation for projectile sizes from crater diameters and other parameters. The primary parameters are projectile density, impact velocity, target density, impact angle, and acceleration of gravity. For Ceres the acceleration of gravity is 0.28 m/sec$^2$. The typical collision velocity for asteroids impacting Ceres is 4.5 – 5.2 km/sec (O'Brien and Sykes, 2011). An impact velocity of 4.85 km/sec was used for both simulations. Impact angles of 45 and 90 degrees were used. The target densities used were for porous rock (1500 kg/m$^3$) and dense rock (3000 kg/m$^3$). The latter density is certainly too high, but it was used to see if there was a significant difference in the shape of the curve. The only difference is a slightly higher position on an R plot. The projectile density for both simulations was 3000 kg m$^{-3}$. However, for comparison with main belt asteroid diameters the vertical position on an R plot is arbitrary since it represents crater density. The different impact parameters do not have any influence on the comparison with asteroid belt diameters.

Figure 9 shows the Ceres projectile diameters for all craters compared to main belt asteroid diameters. The Ceres projectile size distribution is very similar to the main belt asteroids. Since the cratering record on Ceres is almost surely the result of Main Belt Asteroids it follows that the terrestrial planet heavily cratered terrain is consistent with an origin from main belt asteroids.

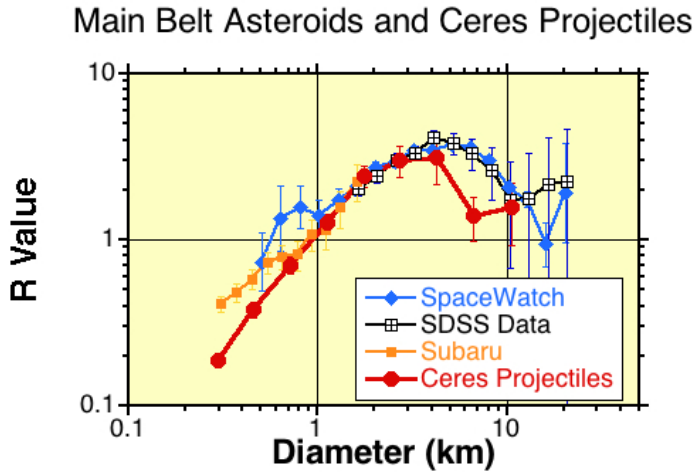

Figure 9. This graph shows the R plots of projectile size distributions of main belt asteroids compared to that derived from all craters on Ceres.

## Conclusions

Ceres has a crater size-frequency distribution very similar to those of heavily cratered surfaces on the terrestrial planets. It also has areas of both relatively high and relatively low crater density, and the crater size-frequency distributions of both are similar to the heavily cratered areas of the terrestrial planets. The lowest crater density is only slightly less than the average crater density (see Figure 5). The heavily cratered terrain on ~15% of the surface has a crater density that is about two times greater than the average density of the heavily cratered terrains on the terrestrial planets. This indicates that the terrestrial planets' heavily cratered surfaces are not saturated. The higher crater density on Ceres is plausibly due to the continuous impact of asteroids from the Main Belt during and after the Late Heavy Bombardment. Since Ceres is in the middle of the main asteroid belt it must have been impacted by Main Belt Asteroids both during and after the Late Heavy Bombardment. About 85% of Ceres has been resurfaced since the end of the LHB. The resurfacing has lowered the overall crater density to almost the same value as the average lunar highlands. The Ceres cratering record is consistent with an impact history that includes the period of Late Heavy Bombardment of the terrestrial planets caused by projectiles originating from Main Belt Asteroids.

**Acknowledgements**

We thank two anonymous reviewers whose comments helped improve this paper. SM acknowledges support from the NASA Dawn project. RM gratefully acknowledges research support from NSF (grant AST-1312498).